\documentclass[11pt,a4paper]{article}

\newif\ifanonymous
\anonymousfalse
\newcommand{\anonauthorname}{Anonymous author}

\usepackage[margin=2.7cm]{geometry}
\usepackage{amsmath,amssymb,amsthm,mathtools}
\usepackage{thm-restate}

\makeatletter
\@ifundefined{newcounteralias}{}{%
  \renewcommand\thmt@autorefsetup{\@xa\def\csname\thmt@envname autorefname\@xa\endcsname\@xa{\thmt@thmname}}%
}
\makeatother

\usepackage{xcolor}
\usepackage{graphicx}
\usepackage{tikz}
\usetikzlibrary{arrows.meta,positioning,calc}
\tikzset{pointvertex/.style={circle,draw,inner sep=1.7pt,minimum size=6mm}}

\usepackage[numbers,sort&compress]{natbib}
\usepackage[hidelinks]{hyperref}
\hypersetup{pdftitle={Tight Bounds for Anonymous Parity Certification},
            pdfauthor={}}
\usepackage{cleveref}

\theoremstyle{plain}
\newtheorem{theorem}{Theorem}
\newtheorem{lemma}[theorem]{Lemma}

\newtheorem*{theorem*}{Theorem}
\theoremstyle{definition}

\newtheorem{open}{Open problem}
\theoremstyle{remark}

\usepackage{fancyhdr}
\title{Tight Bounds for Anonymous Parity Certification}

\ifanonymous
\author{\anonauthorname}
\else
\author{%
  Pedro Montealegre\\[2pt]
  {\small Universidad Adolfo Ibáñez}\\
  {\small Santiago, Chile}\\
  {\small\texttt{p.montealegre@uai.cl}}
  \and
  Jorge Valenzuela\\[2pt]
  {\small Universidad Adolfo Ibáñez}\\
  {\small Santiago, Chile}\\
  {\small\texttt{jorg.valenzuela@alumnos.uai.cl}}}
\fi

\date{}

\definecolor{transferblue}{RGB}{0,0,145}
\definecolor{correctionorange}{RGB}{245, 158, 66}

\tikzset{vertex/.style={circle,draw,fill=white,inner sep=1.7pt,minimum size=6mm},
 ground/.style={vertex,fill=black!5},
 flow/.style={-{Stealth[length=1.5mm]},draw=transferblue},
 flowlabel/.style={fill=white,inner sep=1pt,font=\small},
 interval/.style={vertex,minimum size=4.4mm,draw=transferblue,text=black}}
 
\newcommand{\Z}{\mathbb Z}
\newcommand{\Q}{\mathbb Q}
\newcommand{\ind}{\mathbf 1}
\newcommand{\eps}{\varepsilon}
\newcommand{\Even}{\textnormal{\textsc{Even}}}

\begin{document}
\pagenumbering{gobble} % The cover page is excluded from the 12-page main text.
\maketitle

\begin{abstract}
Local certification is the task of verifying graph properties using certificates assigned to the vertices of a network and checked using only local information. A central question in this area is to understand which properties admit constant-size certificates, independently of the network size. A fundamental test case is the parity of the network order. Bousquet, Feuilloley, Valenzuela, and Zeitoun (DISC 2026) showed that the complexity of certifying even order depends dramatically on the verification model. On arbitrary graphs, three bits per vertex suffice when unique identifiers are available and the verification radius is two. In contrast, in the anonymous radius-one model, they established an $\Omega(\log\log^* n)$ lower bound and an $O(\log n)$ upper bound, leaving a substantial gap in our understanding of the problem. They also showed that constant-size certificates suffice in the anonymous radius-one model on graph classes of bounded expansion.

In this paper, we settle the complexity of anonymous parity
certification on general graphs and further investigate the
role of graph sparsity. Our main contributions are:

\begin{itemize}

\item 
We close the gap left by Bousquet et al., proving that
the local complexity of certifying even order in the
anonymous radius-one model is $\Theta(\log\log n)$.

\item 
We show that on graphs of degeneracy at most $d$,
parity can be certified with $O(\log d)$ bits, while
$\Omega(\log\log d)$ bits are necessary. In particular,
every class of bounded degeneracy admits constant-size
certificates, strictly generalizing the previous result
for bounded-expansion classes.

\item 
We show that constant-size certification cannot be
guaranteed on nowhere-dense classes, a broader notion
of sparsity that includes bounded expansion. Specifically,
we construct a fixed, subgraph-closed nowhere-dense class
on which parity certification still requires
$\Theta(\log\log n)$ bits.

\end{itemize}

Together, our results provide a substantially more complete
picture of anonymous parity certification on sparse graph
classes, showing that its complexity depends not only on
graph sparsity, but on the particular structural notion
of sparsity under consideration.

\end{abstract}

\clearpage
\pagenumbering{arabic}

\section{Introduction}
\label{sec:intro}

Local certification is a distributed mechanism for verifying graph
properties. An all-powerful prover assigns a certificate to every vertex,
and each vertex decides whether to accept by inspecting its local
neighborhood and the certificates it contains. A scheme is correct if
every graph satisfying the property has an assignment accepted
everywhere, whereas on a graph violating it, every assignment is
rejected by at least one vertex. The central resource is the maximum
number of certificate bits per vertex; see~\cite{KKP10,GS16,FKP13,F21}
for background. Recent research has increasingly focused on the
\emph{constant-size} regime: which properties can be certified with
$O(1)$ bits per vertex, independently of the size of the network?
Such properties are sometimes regarded as ``local'' from the
certification perspective, even when their definitions are global
\cite{BFZ25}.

One of the most basic questions in this direction is whether the
number of vertices is even. Let \Even\ denote this property. Parity is
inherently global: changing the number of vertices by one changes the
answer, potentially far away from any given vertex. Yet it can be
remarkably easy to certify. On cycles, even order is equivalent to
bipartiteness and can be certified with one bit per vertex. On trees,
constant-size certificates also suffice, by aggregating subtree sizes
modulo two. The challenge is to understand whether such small local
proofs are possible on arbitrary graphs.

This question was recently investigated by Bousquet, Feuilloley,
Valenzuela, and Zeitoun~\cite{BFVZ} (DISC 2026), who uncovered a striking
dependence on the verification model and the underlying graph class.
On general graphs, they showed that three bits suffice when vertices
have unique identifiers and the verification radius is two. At the
other extreme, they studied the \emph{anonymous radius-one model},
which corresponds to the \emph{MULTISET} model of local verification.
In this model, a vertex sees only its own certificate and the multiset
of its neighbors' certificates, including multiplicities, but has no
identifiers, port numbers, or knowledge of the graph order. They proved an
$\Omega(\log\log^* n)$ lower bound, against an $O(\log n)$ upper bound.
They also established that, on every graph class of \emph{bounded
expansion}, constant-size certificates suffice even in this restricted
anonymous model. Bounded expansion is a robust notion of structural
sparsity that includes, for example, bounded-degree and planar graphs.

Two natural questions remained. First, the gap between
$\Omega(\log\log^* n)$ and $O(\log n)$ left the actual complexity of
anonymous parity certification on general graphs essentially
undetermined. Second, the bounded-expansion result raised a structural
question: how far can constant-size parity certification be extended
beyond bounded expansion, and which weaker notions of sparsity, if any,
still guarantee it? These are the questions we address here.

\subsection{Our results}
\label{sec:results}

%\noindent\textbf{A tight bound on general graphs.}

Our first contribution settles the open problem of~\cite{BFVZ} for
anonymous networks with verification radius one. We establish matching
upper and lower bounds on the size of the certificates.

\begin{restatable}{theorem}{parityUpper}
\label{thm:upper}
There is an anonymous radius-one certification scheme for \Even\ using $O(\log\log n)$ bits per vertex.
\end{restatable}

Our scheme shows that the order of the network need not be
encoded explicitly to certify its parity, even in the anonymous
radius-one model. The natural question is whether the doubly
logarithmic dependence can be reduced further. We show that
it cannot, already on connected bipartite graphs.

\begin{restatable}{theorem}{parityLower}
\label{thm:lower}
For every sufficiently large even $n$, there is a connected bipartite $n$-vertex graph $G_n$ such that every accepting assignment of every correct anonymous radius-one verifier for \Even\ has a certificate of length $\Omega(\log\log n)$. This remains true when correctness is required only on bipartite inputs.
\end{restatable}

Together, Theorems~\ref{thm:upper} and~\ref{thm:lower} show that certifying
\Even\ on connected $n$-vertex graphs has local complexity
$\Theta(\log\log n)$. Thus parity has genuinely intermediate complexity
in this model: constant-size certificates do not suffice, but the
$\Theta(\log n)$ bits needed to explicitly represent the graph order
are unnecessary. The lower bound holds even under a bipartite promise.

\medskip
%\noindent\textbf{From bounded expansion to bounded degeneracy.}
Our second contribution strengthens the positive result of~\cite{BFVZ}
by replacing bounded expansion with bounded degeneracy. Recall that a
graph is $d$-degenerate if every nonempty subgraph has a vertex of
degree at most $d$. This condition allows vertices of arbitrarily
large degree but restricts the density of every subgraph. Crucially,
every bounded-expansion class has bounded degeneracy, whereas the
converse is false: for example, the class of one-subdivisions of
complete graphs is $2$-degenerate but does not have bounded expansion.

\begin{restatable}{theorem}{parityDegeneracyUpper}
\label{thm:degeneracy-upper}
The verifier of Theorem~\ref{thm:upper} admits $O(\log d)$-bit certificates on every connected even-order $d$-degenerate graph, for every integer $d\ge2$, and constant-size certificates when $d=1$. It remains sound on arbitrary connected graphs and is not supplied with $d$.
\end{restatable}

In particular, every graph class of bounded degeneracy admits
constant-size certificates, strengthening the bounded-expansion
result of~\cite{BFVZ}. The dependence on $d$, however, cannot
be removed altogether. Our next result shows that the certificate
size must grow with $d$, even on bipartite graphs and even when
the verifier is only required to work under the $d$-degeneracy
promise.

\begin{restatable}{theorem}{parityDegeneracyLower}
\label{thm:degeneracy-lower}
For every sufficiently large integer $d$, there is a connected bipartite graph of even order and degeneracy at most $d$ on which every accepting assignment of every anonymous radius-one certification scheme for \Even\ contains a certificate of length $\Omega(\log\log d)$. This holds even when correctness is required only on connected bipartite $d$-degenerate graphs.
\end{restatable}

The two bounds leave a gap in the optimal dependence on
the degeneracy parameter, between $\Omega(\log\log d)$
and $O(\log d)$.

\medskip
%\noindent\textbf{The limit of the positive result: nowhere denseness.}

Our third contribution shows that the bounded-expansion theorem cannot
be extended to all nowhere-dense classes. A graph class is
\emph{nowhere dense} if, for every fixed radius $r$, there is a clique
that cannot occur as an $r$-shallow minor of any graph in the class.
This is a broad notion of sparsity: every bounded-expansion class is
nowhere dense, but the converse does not hold, and nowhere-dense
classes may have unbounded degeneracy.

\begin{restatable}{theorem}{parityNowhereDense}
\label{thm:nowhere-dense}
There exists a fixed, subgraph-closed nowhere-dense class
of graphs on which certifying \Even\ in the anonymous
radius-one model has local complexity $\Theta(\log\log n)$.
\end{restatable}

In fact, our construction gives a bipartite class whose girth
grows with the graph order. A precise description of this
class, together with a quantitative lower bound, is given
in Section~\ref{sec:nowhere-dense}. Thus, the full difficulty
of the unrestricted anonymous model persists even on graphs
with arbitrarily large girth.

Taken together, these results give a substantially sharper picture of
parity certification on sparse graph classes. The constant-size result
extends beyond bounded expansion to all classes of bounded degeneracy,
but it fails already on a fixed nowhere-dense class. Hence the positive
phenomenon discovered in~\cite{BFVZ} is not explained by sparsity in
general: the particular structural restriction matters. What remains
open is the exact quantitative cost of parity certification as a
function of degeneracy.

\subsection{Proof ideas and techniques}
\label{sec:overview}

\noindent\textbf{Sharing contributions without identifying parents.}
Earlier anonymous schemes organize contributions by choosing a parent for each vertex. A color identifies that parent only if no other neighbor has the same color~\cite{BFVZ}. Enforcing this uniqueness can require many colors, making the parent descriptions expensive.

Our approach allows a contribution to be shared among several neighbors. For example, two senders with the same two receivers can each send half a unit to each receiver. Each sender sends one unit in total, and each receiver receives one unit, even though neither sender distinguishes between the receivers. Fractions allow us to coordinate whole-number totals without making individual choices.

The verifier checks that the amount sent minus the amount received is an odd integer at every vertex. Every payment appears once as an outgoing amount and once as an incoming amount, so these differences sum to zero. Since an odd number of odd integers cannot sum to zero, acceptance implies that the graph has even order.

The main task is to arrange these payments using short descriptions. Our sharing lemma coordinates fractional payments so that receivers obtain integer totals, using colors and fractions that take $O(\log\log n)$ bits to describe. The prover applies the lemma between successive distance layers, working toward a chosen root. At each step, a vertex chooses what to send according to the parity of what it has already received. On bounded-degeneracy graphs, the same construction uses constant-size descriptions.

\medskip
\noindent\textbf{Changing parity without changing local views.}
A lower bound construction must change the number of vertices; changing edges alone cannot change parity. It must also leave every vertex with a local view that the verifier accepts. The earlier construction~\cite{BFVZ} achieves this using all nonempty subsets of a point set, resulting in exponentially many vertices. We obtain a smaller construction using intervals in an ordered point set. Each interval is represented by a vertex adjacent to the points it contains.

The key is to find two collections of intervals that cover every point equally often, but whose numbers of intervals have different parities. If all their interval vertices carry the same certificate, we can copy one collection and delete the other. Every point loses and gains the same number of neighbors with that certificate. Retained interval vertices keep their views, and new copies imitate existing vertices. All checks still pass, although the graph order changes parity.

Intervals provide a simple way to find such collections. When traversing a closed loop of interval endpoints, rightward and leftward steps cover every point equally often. If the loop has odd length, the two groups of steps have different parities. Thus a sound verifier must prevent an odd cycle whose intervals all carry the same certificate. A classical coloring bound~\cite{HHM77} then forces a logarithmic number of distinct certificates. Because the construction has only polynomially many vertices, describing that many certificates requires $\Omega(\log\log n)$ bits.

\medskip
\noindent\textbf{Preserving large girth through vertex splitting.}
For the nowhere-dense lower bound, we need a parity-changing operation that preserves large girth. Suppose three vertices carry the same certificate and one has a neighbor-certificate profile equal to the sum of the other two profiles. We can split that vertex into two, distributing its neighbors so that each replacement imitates one of the other vertices. Every old neighbor still sees one vertex with the original certificate. The order increases by one, and splitting cannot create a shorter cycle.

To force such a triple, we return to intervals. Their color-count profiles satisfy an exact identity: the profile of a long interval is the sum of the profiles of its two pieces. For every coloring of the point set, a counting argument finds an ordered sequence whose interval profiles are all sufficiently likely under random subset sampling. Sampling supplies many representatives of each profile, allowing us to remove short cycles while keeping all these representatives available. Color each interval's endpoint pair by its representative's certificate. An elementary Ramsey bound gives a monochromatic triangle, and hence the three equally certified profiles needed for splitting.

\subsection{Related work}
\label{sec:related}

\noindent\textbf{Sublogarithmic certification.} Bousquet, Feuilloley, and Zeitoun~\cite{BFZ25} identified a $\Theta(\log\log n)$ regime for an arithmetic property of anonymous path lengths, defined using products of consecutive primes, and proved a gap between constant and logarithmic complexity on anonymous cycles. Our result brings the same intermediate scale to a simple counting predicate on unrestricted connected graphs. Since parity has constant-size certificates on paths and cycles, its difficulty here comes from the richer range of possible topologies. A related result of Feuilloley, Sedl\'a\v{c}ek, and Sl\'avik~\cite{FSS25} gives $O(\log\log n)$-bit certificates for at most one selected vertex when the minimum degree is at least $\alpha\log^2 n$, for a sufficiently large constant $\alpha$. Their scheme uses identifiers and radius two, highlighting the different assumptions under which sublogarithmic certification becomes possible.

\medskip
\noindent\textbf{Structure and certificate size.}
Graph structure is a recurring theme in local certification, from certifying planarity~\cite{Planar21} to logical metatheorems on tree-like graph classes~\cite{BFP22,FMRT24,CKM25}. On bounded-treewidth classes, these metatheorems progressed from $O(\log^2 n)$-bit certificates~\cite{FMRT24} to $O(\log n)$-bit certificates for MSO$_2$ properties~\cite{CKM25}. For parity in the anonymous radius-one model, Bousquet et al.~\cite{BFVZ} obtained constant-size certificates on bounded-expansion classes using Hickingbotham's conflict-free neighborhood colorings~\cite{H23}. We extend this constant-size guarantee from bounded expansion to bounded degeneracy. However, replacing bounded expansion with nowhere denseness does not preserve the guarantee: on a fixed subgraph-closed nowhere-dense class, parity certification still has the $\Theta(\log\log n)$ complexity of general graphs.

\medskip
\noindent\textbf{Combinatorial tools.} Our construction draws on the theory of conflict-free and odd colorings~\cite{ELRS03,PT09,CKP13,DOP24}. Conflict-free colors single out a neighbor, whereas odd colors need only occur an odd number of times in a neighborhood. Rational sharing adds another possibility: contributions can be distributed among indistinguishable neighbors while preserving integral totals. Conservation checks already appear in flow certification~\cite{KKP10,PSP22}; the challenge here is to describe the transfers with short anonymous vertex certificates. The lower-bound techniques also have a broader reach. Chang, Lee, and Tsai~\cite{CLT26} adapted the earlier parity construction to anonymous $2$-vertex-connectivity, showing how neighborhood identities that preserve accepted local views can obstruct certification beyond counting predicates.

\subsection{Discussion and open problems}
\label{sec:discussion}

We have determined the local complexity of \Even\ in the anonymous radius-one model: $\Theta(\log\log n)$ bits are necessary and sufficient in the worst case. Parity thus gives an elementary global property whose optimal certificate size is unbounded but sublogarithmic. The two proofs explain this scale from complementary directions. Rational sharing coordinates parity contributions without selecting unique parents, while the interval construction quantifies how much ambiguity any sound verifier must eliminate. The same verifier admits constant-size certificates on every class of bounded degeneracy. These results settle the dependence on graph order and leave several questions about structure and the verification model.

The $\Theta(\log\log n)$ bound also holds, with a smaller leading constant in the lower bound, on a fixed subgraph-closed nowhere-dense class. Thus nowhere denseness alone does not guarantee constant-size certificates, even though bounded degeneracy does.\\

\noindent\textbf{The dependence on degeneracy.}
Our structural bounds lie between $\Omega(\log\log d)$ and $O(\log(d+1))$. They show that degeneracy alone controls certificate size, but do not determine the optimal dependence on this parameter.

\begin{open}
\label{open:degeneracy}
What is the optimal local complexity of certifying \Even\ on connected $d$-degenerate graphs, as a function of $d$? In particular, do $O(\log\log(d+2))$-bit certificates suffice, independently of the graph order?
\end{open}

Our lower-bound witnesses have order $\Theta(d^2)$, so the general upper bound already gives $O(\log\log d)$-bit certificates on them. Consequently, they cannot establish a larger asymptotic lower bound in terms of $d$. An $\Omega(\log d)$ lower bound would require witnesses whose order exceeds every fixed polynomial in $d$. Within our approach, improving the upper bound would require a more compact description of the sharing rules.\\

\noindent\textbf{Identifiers and verification radius.}
Bousquet et al.~\cite{BFVZ} gave three-bit certificates when identifiers and radius two are both available. Our tight bound establishes the cost of imposing anonymity and radius one together. The intermediate cases, already asked about in~\cite{BFVZ}, remain unresolved.

\begin{open}
\label{open:intermediate-models}
What is the local complexity of certifying \Even\ on arbitrary connected graphs (i) with identifiers and verification radius one, and (ii) anonymously with verification radius two? Do constant-size certificates suffice in either model?
\end{open}

Both models inherit our $O(\log\log n)$ upper bound. The lower-bound modification uses anonymity to copy certified vertices without introducing new identifiers into their local views, and radius one to hide changes in their neighbors' other incident edges. Extending the obstruction to either intermediate model therefore requires an additional idea. Resolving these cases would separate the roles of identifiers and observation radius in the cost of certifying parity.\\

\noindent\textbf{Counting modulo other integers.}
Divisibility by a fixed integer, studied in~\cite{BFVZ}, is a natural next setting in which to test the scope of our techniques.

\begin{open}
\label{open:other-moduli}
For a fixed integer $k\ge3$, what is the optimal certificate size for verifying that the order of a connected graph is divisible by $k$ in the anonymous radius-one model? Does the $\Theta(\log\log n)$ bound hold for every such $k$?
\end{open}

The conservation principle extends directly: if every vertex has integer net outflow congruent to one modulo $k$, then summing the balances forces $k\mid n$. Constructing these balances with short anonymous labels, however, requires control of the residues modulo $k$, beyond the odd--even distinction used here. For the lower bound, the profile-splitting construction also gives an $\Omega(\log\log n)$ lower bound for divisibility by every fixed integer $k\ge2$, since it changes the order by exactly one. The remaining question is whether an $O(\log\log n)$ upper bound holds for these moduli.

\section{Model and definitions}
\label{sec:prelim}

\subsection{Graphs and certification model}
\label{sec:graph-notation}

All graphs are finite, simple, and undirected; input graphs are connected. For $G=(V,E)$, write $n=|V|$, $N_G(v)$ for the neighborhood of $v$, $\deg_G(v)=|N_G(v)|$, $\Delta(G)$ for the maximum degree, and $G[X]$ for the subgraph induced by $X\subseteq V$. We omit the subscript $G$ when clear. All logarithms are base two. Write $\Z$ and $\Q$ for the integers and rationals, and $\ind_A$ for the indicator of a condition or the characteristic vector of a set.

A graph is \emph{$k$-degenerate}, for an integer $k\ge0$, if every nonempty subgraph has a vertex of degree at most $k$. Equivalently, there is an ordering in which each vertex has at most $k$ later neighbors; the least such $k$ is its \emph{degeneracy}. Orienting edges from earlier to later in this ordering gives an acyclic orientation of maximum outdegree at most $k$.

The \emph{girth} of a graph is the length of its shortest cycle, with girth infinity for forests. For an integer $\rho\ge0$, a \emph{depth-$\rho$ minor} of $G$ is obtained from a subgraph of $G$ by contracting pairwise disjoint connected subgraphs of radius at most $\rho$, measured within each contracted subgraph. A graph class $\mathcal C$ is \emph{nowhere dense} if, for every integer $\rho\ge0$, there is an integer $t$ such that no graph in $\mathcal C$ contains $K_t$ as a depth-$\rho$ minor~\cite{NODM12}.

\noindent{\bf Certificates and local verification.}
We use the anonymous model with verification radius one of~\cite{BFVZ}. A \emph{certificate assignment} on $G$ is a map $c:V\to\{0,1\}^*$, chosen by an external prover. The local view at a vertex $v$ is
\[
\operatorname{view}_{G,c}(v)
=\bigl(c(v),\ \{\!\{c(u):u\in N(v)\}\!\}\bigr),
\]
where double braces denote a multiset. A \emph{verification algorithm} is a deterministic function that takes this view and outputs \emph{accept} or \emph{reject}. The same function is used at every vertex and on every input graph. A graph is \emph{globally accepted} if all its vertices accept. Vertices have no identifiers, ports, graph order, neighbor degrees, or information about edges between neighbors or their other incident edges. There are no computational restrictions on the prover or the verification function. In particular, the prover may assign different certificates to structurally indistinguishable vertices.

\noindent{\bf Certification schemes and local complexity.}
Let $\mathcal C$ be a graph class, let $P$ be a graph property on $\mathcal C$, and let $s:\mathbb N\to\mathbb N$. A \emph{certification scheme} for $P$ on $\mathcal C$ with \emph{local complexity} $s$ consists of a verification algorithm satisfying the following two conditions:
\begin{itemize}
\item \emph{Completeness.} Every connected $n$-vertex graph $G\in\mathcal C$ satisfying $P$ admits a certificate assignment $c$ with $|c(v)|\le s(n)$ for every vertex $v$, under which $G$ is globally accepted.
\item \emph{Soundness.} For every connected graph $G\in\mathcal C$ not satisfying $P$, and every certificate assignment $c$, at least one vertex rejects.
\end{itemize}
The class $\mathcal C$ is a promise: the verification algorithm may depend on $\mathcal C$ and $P$, and no correctness requirement is imposed outside $\mathcal C$. Unless a class is specified, $\mathcal C$ consists of all graphs. In this paper, $P=\Even$: completeness concerns even-order graphs and soundness concerns odd-order graphs.

\subsection{Rational transfers, encodings, and standard combinatorial facts}

A transfer on an ordered edge $(v,u)$ contributes to the outgoing total at $v$ and the incoming total at $u$. The \emph{net outflow} $\Delta(v)$ is outgoing minus incoming; an odd net outflow means $\Delta(v)\in2\Z+1$. All transfers used in the upper bounds are nonnegative rationals.

We fix a self-delimiting encoding of nonnegative integers. For $t\ge0$, let $\ell$ be the binary length of $t+1$ and encode $t$ as $1^\ell0\operatorname{bin}(t+1)$, using $2\ell+1=O(\log(t+2))$ bits. Encode a nonnegative rational by its numerator and positive denominator in lowest terms, with zero represented by $0/1$. If $q\in[0,2)$ has reduced denominator at most $k$, its numerator is less than $2k$, so its encoding takes $O(\log(k+1))$ bits.

We use the following standard facts.

\begin{lemma}[Hall~\cite{Hall35}]
\label{lem:hall}
Let $B=(S\cup R,E_B)$ be a finite bipartite graph, and write $N_B(T)=\bigcup_{s\in T}N_B(s)$ for $T\subseteq S$. If
\(|N_B(T)|\ge |T|\text{ for every }T\subseteq S,\)
then $B$ has a matching covering~$S$.
\end{lemma}

\begin{lemma}[Degeneracy bound, Szekeres and Wilf~\cite{SW68}]
\label{lem:degeneracy}
Every finite $d$-degenerate graph has a proper coloring with at most $d+1$ colors.
\end{lemma}

\begin{lemma}[Bipartite covers, Harary, Hsu, and Miller~\cite{HHM77}]
\label{lem:bipartite-cover}
If $r$ bipartite subgraphs cover all edges of $K_t$, then $r\ge\lceil\log t\rceil$.
\end{lemma}

For a hypergraph $\mathcal H$, write $\Delta(\mathcal H)$ for the maximum number of hyperedges containing a vertex, and $\varepsilon(\mathcal H)$ for the minimum hyperedge size. A graph--hypergraph pair $(F,\mathcal H)$ has a common vertex set.

\begin{lemma}[Dai, Ouyang, and Pirot~{\cite[Theorem~5]{DOP24}}]
\label{lem:odd-coloring}
There is a universal constant $C\ge1$ such that the following holds. Let $(F,\mathcal H)$ be a finite graph--hypergraph pair, with $D=\Delta(\mathcal H)\ge2$. If \(\varepsilon(\mathcal H)>C\log D\),
then its vertex set admits a coloring that is proper on $F$ and has some color occurring an odd number of times in each hyperedge of $\mathcal H$, using at most $\Delta(F)+C\log D$ colors.
\end{lemma}

\section{Upper bound in general graphs}
\label{sec:upper}

Earlier anonymous schemes encode parents through conflict-free colors: a vertex selects a color unique among its predecessors~\cite{BFVZ}. This can be expensive even on sparse graphs. If there are $m$ receivers and one sender for every pair, all receivers need different colors. The resulting one-subdivision of $K_m$ has $\Theta(m^2)$ vertices, so color names take $\Theta(\log n)$ bits. Rational sharing avoids selecting a unique parent by allowing equal transfers to neighbors with the same color.

We first construct transfers on a bipartite graph of senders and receivers, then apply this construction between consecutive BFS layers.

\subsection{The sharing lemma}

In a bipartite graph $B=(S\cup R,E_B)$, vertices of $S$ are \emph{senders} and vertices of $R$ are \emph{receivers}. A receiver coloring is \emph{odd at a sender} if some color occurs an odd number of times among its neighbors. Each sender stores a base amount and a selected color; each receiver stores its own color. These labels determine the edge transfers, including when senders have identical neighborhoods.

\begin{lemma}[Sharing lemma]
\label{lem:sharing}
Let $B=(S\cup R,E_B)$ be a finite bipartite graph on $n\ge2$ vertices, with senders $S$ and receivers $R$. Assume every sender has a neighbor. There are receiver colors $\gamma_r\in\Z_{>0}^2$, rational base amounts $q_s\in[0,2)$, and selected colors $a_s\in\{0\}\cup\Z_{>0}^2$ such that the transfers
\[
f(s,r)=q_s+\ind_{\gamma_r=a_s}\qquad(sr\in E_B)
\]
satisfy
\begin{equation}
\begin{aligned}
\sum_{r\in N_B(s)}f(s,r)&\in2\Z+1 &&(s\in S),\\
\sum_{s\in N_B(r)}f(s,r)&\in\Z &&(r\in R).
\end{aligned}
\label{eq:sharing}
\end{equation}
There are $O(\log^2(n+2))$ colors, each coordinate of a color is $O(\log(n+2))$, and every base amount has reduced denominator $O(\log(n+2))$. Each label $\gamma_r$, $q_s$, or $a_s$ therefore has an $O(\log\log(n+2))$-bit encoding, with color coordinates encoded separately. The value $a_s=0$ selects no color; equality of colors means equality in both coordinates, and no receiver color equals $0$.
\end{lemma}

\begin{proof}
Let $C$ be the constant in Lemma~\ref{lem:odd-coloring}. Take the least positive even integer $b>C\log(n+2)$, and put $k=2b-1$. Partition the senders into
\[
S_{\mathrm{small}}=\{s\in S:\deg_B(s)\le k\},
\qquad
S_{\mathrm{large}}=\{s\in S:\deg_B(s)\ge k+1\}.
\]
We construct transfer rules for these two groups and then combine their receiver colorings.

\noindent{\bf Case 1: senders with at most $k$ neighbors.}
For each $s\in S_{\mathrm{small}}$, write $d_s=\deg_B(s)$. Choose rational base amounts $q_s$ such that
\[
d_sq_s\in\Z\quad(s\in S_{\mathrm{small}}),
\qquad
\sum_{s\in N_B(r)\cap S_{\mathrm{small}}}q_s\in\Z\quad(r\in R).
\]
Thus both outgoing and incoming base totals are integral. The zero assignment is feasible. Among feasible assignments, maximize the number of senders with odd outgoing base total. Call these senders \emph{satisfied}, and the others \emph{unsatisfied}. A maximum exists because this number is an integer between zero and $|S_{\mathrm{small}}|$.

We claim that the unsatisfied senders can be matched to distinct receivers. Otherwise Hall's theorem gives a set $T$ of unsatisfied senders with $|N_B(T)|<|T|$. Their neighborhood vectors are supported on fewer coordinates than there are vectors, so they are linearly dependent over $\Q$. Clearing denominators gives integers $t_s$, not all zero, such that
\[
\sum_{s\in T}t_s\ind_{N_B(s)}=0.
\]
Let $g$ be the positive greatest common divisor of the nonzero integers $d_st_s$. Replace $q_s$ by $q_s+t_s/g$ for $s\in T$. At a receiver $r$, the incoming total changes by
\(
\frac1g\sum_{s\in T\cap N_B(r)}t_s=0
\).
At sender $s\in T$, the outgoing total changes by the integer $d_st_s/g$. These integers have greatest common divisor one, so at least one is odd. At least one previously even total becomes odd, and all previously satisfied senders are unchanged. This contradicts maximality. Hence there is a matching covering the unsatisfied senders; let $\rho(s)\in N_B(s)$ be the receiver matched to $s$. Figure~\ref{fig:sharing-perturbation} illustrates how a dependence preserves receiver totals while improving sender parities.

\begin{figure}[t]
\centering
\begin{minipage}[t]{.47\linewidth}
\vspace{0pt}\centering
\textbf{\small(a) Three unsatisfied senders}\par\smallskip
\begin{tikzpicture}[x=1cm,y=1cm,font=\small]
 \path[use as bounding box] (-.45,-.7) rectangle (3.95,2.8);
 \node[vertex] (a) at (0,1.65) {$a$};
 \node[vertex] (b) at (1.75,1.65) {$b$};
 \node[vertex] (c) at (3.5,1.65) {$c$};
 \node[pointvertex] (u) at (.75,0) {$u$};
 \node[pointvertex] (v) at (2.75,0) {$v$};
 \draw[flow,draw=black] (a)--(u);
 \draw[flow,draw=black] (a)--(v);
 \draw[flow,draw=black] (b)--(u);
 \draw[flow,draw=black] (b)--(v);
 \draw[flow,draw=black,densely dashed] (c)--(u);
 \draw[flow,draw=black,densely dashed] (c)--(v);
 \node[anchor=south,align=center,font=\footnotesize] at (0,2.04)
   {$q_a=1$\\sends $2$};
 \node[anchor=south,align=center,font=\footnotesize] at (1.75,2.04)
   {$q_b=1$\\sends $2$};
 \node[anchor=south,align=center,font=\footnotesize] at (3.5,2.04)
   {$q_c=0$\\sends $0$};
 \node[anchor=north,font=\footnotesize] at (.75,-.38) {receives $2$};
 \node[anchor=north,font=\footnotesize] at (2.75,-.38) {receives $2$};
\end{tikzpicture}
\end{minipage}\hfill
\begin{minipage}[t]{.47\linewidth}
\vspace{0pt}\centering
\textbf{\small(b) Two become satisfied}\par\smallskip
\begin{tikzpicture}[x=1cm,y=1cm,font=\small]
 \path[use as bounding box] (-.45,-.7) rectangle (3.95,2.8);
 \node[vertex,draw=transferblue, very thick] (a) at (0,1.65) {$a$};
 \node[vertex,draw=transferblue, very thick] (b) at (1.75,1.65) {$b$};
 \node[vertex] (c) at (3.5,1.65) {$c$};
 \node[pointvertex] (u) at (.75,0) {$u$};
 \node[pointvertex] (v) at (2.75,0) {$v$};
 \draw[flow,draw=black] (a)--(u);
 \draw[flow,draw=black] (a)--(v);
 \draw[flow,draw=black] (b)--(u);
 \draw[flow,draw=black] (b)--(v);
 \draw[flow,draw=black,densely dashed] (c)--(u);
 \draw[flow,draw=black,densely dashed] (c)--(v);
 \node[anchor=south,align=center,font=\footnotesize] at (0,2.04)
   {$q_a=\tfrac12$\\sends $\mathbf{1}$};
 \node[anchor=south,align=center,font=\footnotesize] at (1.75,2.04)
   {$q_b=\tfrac32$\\sends $\mathbf{3}$};
 \node[anchor=south,align=center,font=\footnotesize] at (3.5,2.04)
   {$q_c=0$\\sends $0$};
 \node[anchor=north,font=\footnotesize] at (.75,-.38) {receives $2$};
 \node[anchor=north,font=\footnotesize] at (2.75,-.38) {receives $2$};
\end{tikzpicture}
\end{minipage}
\smallskip

\caption{A rational perturbation in the maximality argument. All three senders $a,b,c$ have neighborhood $\{u,v\}$, so the unsatisfied set in (a) violates Hall's condition. Its neighborhood vectors admit the dependence $t=(-1,1,0)$; since every sender has degree two, $g=2$. Adding $t/g$ to the base amounts gives (b): each receiver's incoming total is unchanged, but two senders become satisfied as their outgoing totals change from even to odd. Thus (a) cannot maximize the number of satisfied senders.}
\label{fig:sharing-perturbation}
\end{figure}

Subtract an even integer from each $q_s$ to put it in $[0,2)$. All incoming and outgoing totals change by even integers, preserving integrality and parity. Since $d_sq_s$ is integral, the reduced denominator of $q_s$ divides $d_s\le k$, and its numerator is less than $2d_s$. Thus the base amount takes $O(\log(k+1))$ bits.

To make a unit correction at each unsatisfied sender, form an auxiliary graph $K_{\mathrm{small}}$ on the receivers. For every such sender $s$, join $\rho(s)$ to all other vertices of $N_B(s)$. For $X\subseteq R$, every edge of $K_{\mathrm{small}}[X]$ is contributed by a matched receiver in $X$. These receivers are distinct and each contributes at most $k-1$ edges, so
\[
|E(K_{\mathrm{small}}[X])|\le(k-1)|X|.
\]
The graph $K_{\mathrm{small}}$ is therefore $(2k-2)$-degenerate. Lemma~\ref{lem:degeneracy} gives a proper coloring $\chi$ with $2k-1$ colors. Each matched receiver has a color unique among its sender's neighbors. Select this color at an unsatisfied sender, adding exactly one to its outgoing total. Select no color at a satisfied sender. Every sender in this case now sends an odd integer total, and receiver totals remain integral.

\noindent{\bf Case 2: senders with at least $2b$ neighbors.}
Let $\mathcal H$ be the hypergraph on $R$ whose hyperedges are the distinct neighborhoods $N_B(s)$, for $s\in S_{\mathrm{large}}$. Set $D=\Delta(\mathcal H)$, taking $D=0$ if there are no hyperedges. A receiver belongs to at most one hyperedge per large sender, so \(D\le |S_{\mathrm{large}}|\le n\).

If $D\le1$, the distinct neighborhoods are pairwise disjoint. Give one receiver in each such neighborhood color one and all remaining receivers color two. This is odd at every large sender; if there are no large senders, any coloring suffices.

If $D\ge2$, apply Lemma~\ref{lem:odd-coloring} to the pair $(F,\mathcal H)$, where $F=(R,\varnothing)$ is the edgeless graph on the receivers. The parameters in the lemma satisfy
\[
\begin{aligned}
\Delta(F)&=0,\\
\varepsilon(\mathcal H)
&=\min_{s\in S_{\mathrm{large}}}\deg_B(s)\ge2b\\
&>C\log(n+2)>C\log D,
\end{aligned}
\]
and the lemma supplies an odd coloring of $\mathcal H$ with at most $C\log D<b$ colors. Allowing unused colors, both cases therefore give a coloring $h:R\to\{1,\ldots,2b\}$ that is odd at every sender in $S_{\mathrm{large}}$: each sender's neighborhood is a hyperedge, including when several senders have the same neighbors.

\noindent{\bf Combining the receiver colorings.}
Give each receiver $r \in R$ the ordered-pair color $\gamma_r=(h(r),\chi(r))$. This is a common refinement of the two colorings: two receivers have the same color precisely when they agree in both coordinates. There are at most $2b(2k-1)=O(\log^2(n+2))$ color pairs. For a satisfied sender in $S_{\mathrm{small}}$, retain $q_s$ and set $a_s=0$. For an unsatisfied sender, retain $q_s$ and select the refined color of $\rho(s)$. This receiver still has a color unique in $N_B(s)$, so all transfers from low-degree senders are unchanged.

For a sender in $S_{\mathrm{large}}$, an odd $h$-color class in its neighborhood splits into refined color classes. At least one remains odd. Set $q_s=0$ and select that refined color as $a_s$. This sends an odd integer total and contributes only integers at receivers. The low-degree senders already gave integral receiver totals, proving~\eqref{eq:sharing}.

Every reduced denominator is at most $k=O(\log(n+2))$; its numerator is less than $2k$ because $0\le q_s<2$. The two color coordinates are at most $2b$ and $2k-1$, respectively, both $O(\log(n+2))$. Encoding the coordinates separately, and using a constant-size tag to distinguish a selected pair from $0$, gives $O(\log\log(n+2))$ bits for each color label. The rational bounds give the same bound for each base amount.
\end{proof}

\subsection{The certification scheme}

\parityUpper*

\begin{proof}

\noindent{\bf Certificate and verifier.}
A certificate is a four-tuple \((\lambda,\gamma,q,a)\), where $\lambda\in\{0,1,2\}$ is a layer label, $\gamma\in\Z_{>0}^2$ is a color, $q\in\Q\cap[0,2)$ is a base amount, and $a\in\{0\}\cup\Z_{>0}^2$ is a selected color. Encode $\lambda$ in two bits. Encode $\gamma$ by concatenating the self-delimiting encodings of its two positive coordinates, using the integer encoding from the preliminaries. Record $q$ by its numerator and positive denominator in lowest terms, using the same integer encoding. Encode $a$ with a tag bit: $0$ represents no selected color, and $1$ is followed by the two encoded coordinates of the selected color. A certificate is \emph{malformed} if its fields cannot be decoded, violate the stated ranges, record a fraction that is not in lowest terms, or leave unused trailing bits. Color equality is coordinatewise. Since no receiver color is $0$, the selected color $a=0$ always means that no extra unit is sent.

For an arbitrary certificate assignment, call a neighbor $u$ of $v$ a predecessor when $\lambda_u\equiv\lambda_v-1\pmod3$, and a successor when $\lambda_u\equiv\lambda_v+1\pmod3$. Use subscripts to indicate the vertex supplying a field and write
\(
p_v(u)=q_v+\ind_{\gamma_u=a_v}
\).
A vertex rejects if any certificate in its local view is malformed. Otherwise it checks
\begin{equation}
\Delta(v):=
\underbrace{\sum_{\substack{u\in N(v)\\\lambda_u\equiv\lambda_v-1\pmod3}}p_v(u)}
_{\text{sent to predecessors}}
-
\underbrace{\sum_{\substack{u\in N(v)\\\lambda_u\equiv\lambda_v+1\pmod3}}p_u(v)}
_{\text{received from successors}}
\in2\Z+1.
\label{eq:balance}
\end{equation}
The transfers are nonnegative rationals. The vertex computes its net outflow exactly and checks that it is an odd integer; a noninteger difference causes rejection. Both sums are computable from the local view: a neighbor's fields determine its term, and repeated certificates contribute with their multiplicity. Three residues distinguish the two directions along consecutive layers; only these constant-size residues are stored, rather than full distances.

\noindent{\bf Soundness.}
Each transfer is counted once positively at its sender and once negatively at its receiver; edges between equal residues are ignored. This holds for arbitrary certificate assignments, regardless of whether the residues describe actual layers. If every vertex accepts, conservation gives
\[
0=\sum_{v\in V}\Delta(v)\equiv n\pmod2,
\]
because each summand is an odd integer. Thus $n$ is even.

\noindent{\bf Completeness.}
Choose a root $r$ in an even-order graph and let
\[
L_i=\{v:\operatorname{dist}(r,v)=i\}\qquad(i\ge0),
\quad L_{-1}=\varnothing,
\]
be its BFS layers. Give each vertex its layer number modulo three. Every edge joins equal or consecutive layers, and every nonroot vertex has a neighbor in the preceding layer.

Process $L_h,L_{h-1},\ldots,L_1$, where $h$ is the maximum distance from the root; the root will be handled last. Once a vertex $v\in L_i$ has received an integer amount $S(v) =\sum_{w\in N(v)\cap L_{i+1}}p_w(v)$ from the next layer,
its outgoing total must have parity opposite to $S(v)$. If $S(v)$ is odd, set $q_v=a_v=0$: the vertex sends zero, so its net outflow is odd. Call vertices with even $S(v)$ active, and let $S_i$ be the active vertices of $L_i$. 

Form the bipartite graph $B_i$ with parts $S_i$ and $L_{i-1}$, containing all edges of $G$ between these parts. Its senders are the active vertices and its receivers are the vertices of the preceding layer. It has at most $n$ vertices, and every sender has a receiver neighbor. If $S_i\ne\varnothing$, apply Lemma~\ref{lem:sharing} to $B_i$: store its receiver colors on $L_{i-1}$ and its base amounts and selected colors on $S_i$. On an edge $vu$ of $B_i$, its transfer $f(v,u)$ is exactly the amount $p_v(u)$ computed by the verifier. Every active vertex sends an odd integer total, and every vertex of $L_{i-1}$ receives an integer amount. This continues the induction, starting with zero incoming totals at the deepest layer. If $S_i=\varnothing$, all transfers from that layer are zero and all vertices of the preceding layer may be given color $(1,1)$. Any color not assigned by this procedure may be set to $(1,1)$.

The color of a vertex is chosen when the next layer is processed; its base amount and selected color are chosen when its own layer is processed. Every nonroot vertex now has an odd integer net outflow: active vertices send an odd total after receiving an even total, and inactive vertices send zero after receiving an odd total. Give the root $q=a=0$. All other checks already hold, so conservation forces its net outflow to be \(\Delta(r)=-\sum_{v\ne r}\Delta(v)\).
This is an odd integer because $n-1$ is odd. This gives an accepting assignment.

\noindent{\bf Certificate size.}
By the sharing lemma, each color, base amount, and selected color can be encoded in $O(\log\log n)$ bits. The layer label takes a constant number of bits. A fixed self-delimiting encoding preserves this asymptotic bound. Thus an anonymous radius-one verifier uses $O(\log\log n)$ certificate bits. Figure~\ref{fig:sharing-routing} illustrates the construction, including fractional transfers and a unit correction.
\end{proof}

\begin{figure}[t]
\centering
\begin{tikzpicture}[x=1cm,y=1cm,font=\small]
 \node[anchor=east] at (-4.25,2.9) {$L_2$};
 \node[anchor=east] at (-4.25,1.25) {$L_1$};
 \node[anchor=east] at (-4.25,-.35) {$L_0$};
 \node[vertex] (a) at (-3.2,2.9) {$a$};
 \node[vertex] (b) at (0,2.9) {$b$};
 \node[vertex] (c) at (3.2,2.9) {$c$};
 \node[pointvertex] (u) at (-1.9,1.25) {$u$};
 \node[pointvertex] (v) at (1.9,1.25) {$v$};
 \node[pointvertex] (r) at (0,-.35) {$r$};
 \draw[flow,draw=black] (a)--node[flowlabel,pos=.37,below left] {$\tfrac12$}(u);
 \draw[flow,draw=black] (a)--node[flowlabel,pos=.22,above] {$\tfrac12$}(v);
 \draw[flow,draw=black] (b)--node[flowlabel,pos=.15,left] {$\tfrac32$}(u);
 \draw[flow,draw=black] (b)--node[flowlabel,pos=.15,right] {$\tfrac32$}(v);
 \draw[flow,draw=correctionorange,very thick] (c)--node[flowlabel,pos=.22,above,text=correctionorange] {$1$}(u);
 \draw[flow,draw=black, densely dashed] (c)--
   node[flowlabel,pos=.37,below right] {$0$}(v);
 \draw[flow,draw=black, densely dashed] (u)--
   node[flowlabel,below left] {$0$}(r);
 \draw[flow,draw=black] (v)--node[flowlabel,below right] {$1$}(r);
 \node[anchor=south,align=center,font=\footnotesize] at (-3.2,3.25)
   {$q=\tfrac12$\\no selected color};
 \node[anchor=south,align=center,font=\footnotesize] at (0,3.25)
   {$q=\tfrac32$\\no selected color};
 \node[anchor=south,align=center,font=\footnotesize] at (3.2,3.25)
   {$q=0$\\select $\gamma_u$};
 \node[anchor=east,align=right,font=\footnotesize] at (-2.3,1.15)
   {$\gamma_u=(1,1)$\\receives $3$};
 \node[anchor=west,align=left,font=\footnotesize] at (2.3,1.15)
   {$\gamma_v=(1,2)$\\receives $2$};
 \node[anchor=north,font=\footnotesize] at (0,-.72)
   {$\bigl(\Delta(a),\Delta(b),\Delta(c),\Delta(u),\Delta(v),\Delta(r)\bigr)
      =(1,3,1,-3,-1,-1)$};
\end{tikzpicture}
\caption{Rational sharing on a six-vertex graph. In the first sharing step, the top vertices are senders and $u,v$ are receivers; each sender is adjacent to both receivers. The orange edge adds one unit to the selected color; dashed edges carry zero. The integer receipts determine which receivers become active senders to the preceding layer. Every net outflow is odd, and their sum is zero.}
\label{fig:sharing-routing}
\end{figure}

\section{Lower bound}
\label{sec:lower}

The earlier lower bound~\cite{BFVZ} uses all nonempty subsets of an $m$-point set, giving exponentially many vertices, and a finite-unions argument. We use only intervals, giving quadratically many vertices. An odd cycle of interval endpoints yields a parity-changing modification that preserves local views, and a bipartite-cover count then gives the quantitative certificate bound.

\subsection{The interval incidence construction}

For $m\ge1$, take the ordered set of points $P=\{1,\ldots,m\}$. The graph $H_m$ has a point vertex $x_t$ for each $t\in P$ and an interval vertex $y_{ij}$ for every $0\le i<j\le m$, adjacent exactly to \(\{x_{i+1},\ldots,x_j\}\).
The corresponding subset of $P$ is the discrete interval $(i,j]=\{i+1,\ldots,j\}$. The endpoint indices $0,\ldots,m$ specify the boundaries of these intervals; they do not add vertices to $H_m$. The graph $H_m$ is the bipartite \emph{incidence graph} of these intervals: the two parts consist of point vertices and interval vertices. This graph is connected and bipartite, with \(N_m=m+\binom{m+1}{2}=\frac{m(m+3)}2\) vertices. When $m\equiv1\pmod4$, its order is even. Indeed, $y_{0m}$ is adjacent to every point vertex, and every other interval vertex has a nonempty set of point neighbors, proving connectedness.

Fix a sufficiently large even $n$. Choose the largest $m\equiv1\pmod4$ with $N_m\le n$, and attach $n-N_m$ leaves to the point vertex $x_1$. Call the resulting graph $G_n$. It is connected and bipartite of order $n$. Moreover $N_{m+4}-N_m=4m+14$, so $n=N_m+O(m)$ and $m=\Theta(\sqrt n)$. These additional leaves give a graph for every sufficiently large even $n$.

Consider any accepting certificate assignment on $G_n$. Introduce a separate complete graph $K_{m+1}$ with vertex set $\{0,\ldots,m\}$, and color its edge $ij$ by the certificate of $y_{ij}$. This auxiliary graph is not part of the input graph $G_n$. Its vertices are the interval endpoints, whereas the two parts of $H_m$ represent points and intervals. The input graph $G_n$ remains bipartite even when the auxiliary graph contains odd cycles.

\subsection{A modification preserving local views}

The following lemma is the key local-view argument. A cycle of endpoints gives two collections of intervals that cover each point equally often, allowing additions and deletions of interval vertices to cancel at every point vertex.

\begin{lemma}
\label{lem:bipartite}
For every certificate $b$, the subgraph of the auxiliary graph formed by the edges of color $b$ is bipartite.
\end{lemma}

\begin{proof}
Suppose that it contains a simple odd cycle $t_0,t_1,\ldots,t_\ell=t_0$. For its $a$th edge, let $\eps_a=1$ if $t_a>t_{a-1}$ and $\eps_a=-1$ if $t_a<t_{a-1}$, and write
\[
I_a=(\min(t_{a-1},t_a),\max(t_{a-1},t_a)].
\]
The interval vertices corresponding to these edges all have certificate $b$. For $0\le t\le m$, let $p_t=\ind_{\{1,\ldots,t\}}\in\Z^m$; in particular, $p_0=0$. These vectors record the points at most $t$. By the definition of $I_a$ and $\eps_a$ it holds that \(\eps_a\ind_{I_a}=p_{t_a}-p_{t_{a-1}}.\)
Consequently
\begin{equation}
\sum_{a=1}^{\ell}\eps_a\ind_{I_a}=0.
\label{eq:interval}
\end{equation}
The differences sum to zero because each vector $p_{t_a}$ appears once with each sign as we go around the cycle. Call an interval positive if $\eps_a=1$ and negative if $\eps_a=-1$. Equation~\eqref{eq:interval} says that each point is covered equally often by positive and negative intervals, even though their total number is odd. Figure~\ref{fig:surgery} gives a five-cycle example.

Modify the graph and its certificate assignment as follows. For every $\eps_a=-1$, delete the corresponding interval vertex. For every $\eps_a=1$, add one new copy of that interval vertex, with certificate $b$ and the same point neighbors. Cycle edges are distinct, so the modifications are well defined. The resulting graph is still simple and bipartite: the copies are new vertices adjacent only to point vertices, even when their neighbor sets coincide with those of existing vertices.

At each point vertex, the change in the number of neighbors with certificate $b$ is zero by \eqref{eq:interval}. All other neighbor-certificate multiplicities are unchanged. Its local view is therefore unchanged. Every retained interval vertex and every leaf added when constructing $G_n$ keeps its original local view. Each copy has the same local view as the vertex it copies. All vertices therefore accept.

The order changes by $\sum_a\eps_a$, which is odd because $\ell$ is odd. The modified graph thus has odd order. Some connected component has odd order and all its vertices still accept: taking a component changes no local view, and the verifier is not given the graph order. This contradicts soundness. There is therefore no odd cycle in the subgraph of color $b$, so it is bipartite.
\end{proof}

\begin{figure}[t]
\centering
\begin{minipage}[t]{.28\linewidth}
\vspace{0pt}\centering
\textbf{(a) Auxiliary cycle}\par\smallskip
\begin{tikzpicture}[scale=.77,font=\small,
 cyclelabel/.style={font=\scriptsize,inner sep=.4pt}]
 \node[vertex] (t0) at (0,1.45) {$0$};
 \node[vertex] (t1) at (1.38,.45) {$1$};
 \node[vertex] (t3) at (.85,-1.17) {$3$};
 \node[vertex] (t4) at (-.85,-1.17) {$4$};
 \node[vertex] (t2) at (-1.38,.45) {$2$};
 \draw[flow] (t0)--node[cyclelabel,above right=2pt] {$+A$}(t1);
 \draw[flow] (t1)--node[cyclelabel,right=3pt] {$+B$}(t3);
 \draw[flow] (t3)--node[cyclelabel,below=3pt] {$+C$}(t4);
 \draw[flow,draw=correctionorange] (t4)--
    node[cyclelabel,left=3pt,text=correctionorange] {$-D$}(t2);
 \draw[flow,draw=correctionorange] (t2)--
    node[cyclelabel,above left=2pt,text=correctionorange] {$-E$}(t0);
\end{tikzpicture}
\end{minipage}\hfill
\begin{minipage}[t]{.32\linewidth}
\vspace{0pt}\centering
\textbf{(b) Before}\par\smallskip
\begin{tikzpicture}[x=.85cm,y=.85cm,font=\small]
 \node[pointvertex] (x1) at (0,0) {$x_1$};
 \node[pointvertex] (x2) at (0,1.15) {$x_2$};
 \node[pointvertex] (x3) at (0,2.3) {$x_3$};
 \node[pointvertex] (x4) at (0,3.45) {$x_4$};
 \node[interval,label=right:$A$] (A) at (2,0) {$b$};
 \node[interval,draw=correctionorange,label=right:$E$] (E) at (2,.86) {$b$};
 \node[interval,label=right:$B$] (B) at (2,1.72) {$b$};
 \node[interval,draw=correctionorange,label=right:$D$] (D) at (2,2.59) {$b$};
 \node[interval,label=right:$C$] (C) at (2,3.45) {$b$};
 \draw[transferblue] (x1)--(A) (x2)--(B) (x3)--(B) (x4)--(C);
 \draw[correctionorange] (x1)--(E) (x2)--(E) (x3)--(D) (x4)--(D);
\end{tikzpicture}
\end{minipage}\hfill
\begin{minipage}[t]{.34\linewidth}
\vspace{0pt}\centering
\textbf{(c) After}\par\smallskip
\begin{tikzpicture}[x=.85cm,y=.85cm,font=\small]
 \node[pointvertex] (x1) at (0,0) {$x_1$};
 \node[pointvertex] (x2) at (0,1.15) {$x_2$};
 \node[pointvertex] (x3) at (0,2.3) {$x_3$};
 \node[pointvertex] (x4) at (0,3.45) {$x_4$};
 \node[interval,label=right:$A$] (A) at (2,-.23) {$b$};
 \node[interval,dashed,label=right:$A'$] (AA) at (2,.33) {$b$};
 \node[interval,label=right:$B$] (B) at (2,1.45) {$b$};
 \node[interval,dashed,label=right:$B'$] (BB) at (2,2.01) {$b$};
 \node[interval,label=right:$C$] (C) at (2,3.12) {$b$};
 \node[interval,dashed,label=right:$C'$] (CC) at (2,3.68) {$b$};
 \draw[transferblue] (x1)--(A) (x2)--(B) (x3)--(B) (x4)--(C);
 \draw[transferblue,dashed] (x1)--(AA) (x2)--(BB) (x3)--(BB) (x4)--(CC);
\end{tikzpicture}
\end{minipage}
\smallskip
\begin{center}
\(
A=(0,1],\quad B=(1,3],\quad C=(3,4],\quad
D=(2,4],\quad E=(0,2],\)\\
\(\ind_A+\ind_B+\ind_C=\ind_D+\ind_E.\)
\end{center}
\caption{An odd cycle gives a parity-changing modification. All five interval vertices have certificate $b$. The traversal $0,1,3,4,2,0$ makes $A,B,C$ positive and $D,E$ negative. Delete $D,E$ and add the dashed copies $A',B',C'$. Each point vertex loses and gains one neighbor with certificate $b$, while the order increases by one. Unshown vertices and edges are unchanged.}
\label{fig:surgery}
\end{figure}

\subsection{Counting certificates}

\parityLower*

\begin{proof}
Let $r$ be the number of distinct certificates on interval vertices. By the preceding argument, their endpoint color classes are bipartite subgraphs whose edge sets partition $K_{m+1}$. Lemma~\ref{lem:bipartite-cover} therefore gives \(r\ge\lceil\log(m+1)\rceil\).
This invokes the classical counting bound; the local-view modification above is what makes it applicable to certificates. If all certificates in the assignment have length at most $s$, then $r\le\sum_{j=0}^{s}2^j<2^{s+1}$. Since $m=\Theta(\sqrt n)$, this yields
\[
s\ge\log\log(m+1)-1=\Omega(\log\log n).\qedhere
\]
\end{proof}

\section{Bounds parameterized by degeneracy}
\label{sec:degeneracy}

We now prove bounds for certifying \Even\ on $d$-degenerate graphs. In the general construction, the description of a transfer depends on the order of the bipartite graph used at each layer. We replace this dependence by degeneracy. The key is that every graph $B_i$ between active vertices and their predecessors is a subgraph of the input graph. Its degeneracy is thus at most $d$, even though a sender may have arbitrarily many receiver neighbors. The certificate format, BFS induction, and conservation check remain unchanged. For the lower bound, we check that the interval modification preserves the promised class.

\subsection{Sharing on a \texorpdfstring{$d$}{d}-degenerate bipartite graph}

The low-degree case of Lemma~\ref{lem:sharing} already works with any prescribed degree bound. For senders of higher degree, degeneracy gives a receiver coloring with a unique color in each neighborhood, replacing the odd-coloring theorem used in the general proof.

\begin{lemma}
\label{lem:degenerate-sharing}
Let $d\ge1$ be an integer and let $B=(S\cup R,E_B)$ be a finite $d$-degenerate bipartite graph, with senders $S$ and receivers $R$. Assume every sender has a neighbor, and put $M_d=2d^2+6d-1$.
There are receiver colors $\gamma_r\in\{1,\ldots,M_d\}\times\{1\}$, base amounts $q_s\in\Q\cap[0,2)$ with reduced denominator at most $2d$, and selected colors $a_s\in\{0\}\cup(\{1,\ldots,M_d\}\times\{1\})$ such that
\[
f(s,r)=q_s+\ind_{\gamma_r=a_s}\qquad(sr\in E_B)
\]
satisfies~\eqref{eq:sharing}: every sender sends an odd integer total, and every receiver receives an integer total. Each label has an $O(\log(d+1))$-bit encoding.
\end{lemma}

\begin{proof}
Partition the senders by degree:
\[
S_{\mathrm{small}}=\{s\in S:\deg_B(s)\le2d\},
\qquad
S_{\mathrm{large}}=\{s\in S:\deg_B(s)\ge2d+1\}.
\]

\noindent{\bf Low-degree senders: reuse the rational construction.}
Apply the maximality and matching argument from Case~1 in the proof of Lemma~\ref{lem:sharing}, with degree bound $2d$, to $B[S_{\mathrm{small}}\cup R]$. It gives rational base amounts $q_s\in[0,2)$ with integer outgoing totals $\deg_B(s)q_s$ and integer incoming totals at all receivers. The reduced denominator divides $\deg_B(s)$ and is therefore at most $2d$. The unsatisfied senders, whose outgoing base totals are even, have a matching to distinct receivers $\rho(s)\in N_B(s)$.

Keep the auxiliary graph $K_{\mathrm{small}}$ joining each matched receiver $\rho(s)$ to the other neighbors of $s$, and postpone its coloring. The same counting argument gives
\begin{equation}
|E(K_{\mathrm{small}}[X])|\le(2d-1)|X|\qquad(X\subseteq R).
\label{eq:small-degenerate-density}
\end{equation}
This part of the proof changes only the degree threshold.

\noindent{\bf High-degree senders: use degeneracy.}
Orient the whole graph $B$ in a degeneracy ordering, so every vertex has outdegree at most $d$. This is an auxiliary orientation for choosing the colors; transfers will still go from senders to receivers, regardless of these directions. For $s\in S_{\mathrm{large}}$, let $O_s=\{r\in N_B(s):r\to s\}$. Given $B$ is $d$-degenerate, at most $d$ edges leave $s$ in the auxiliary orientation, so at most $d$ of its receivers are outside $O_s$. In particular, \(|N_B(s)\setminus O_s|\le d\) and since $\deg_B(s)\ge2d+1$, it follows that $|O_s|\ge d+1$. Each receiver belongs to at most $d$ of the neighborhoods $O_s$, since each membership uses an edge leaving that receiver. These are the two consequences of degeneracy used in the construction: few exceptions at a sender, and bounded participation at a receiver. Figure~\ref{fig:degenerate-candidates} illustrates how the first bound ensures that a candidate color does not recur outside $O_s$.

For each $s\in S_{\mathrm{large}}$, arbitrarily choose  $d+1$ candidate receivers $T_s\subseteq O_s$. Candidates may be shared by different senders. Form a graph $K_{\mathrm{large}}$ on $R$ by joining each candidate in $T_s$ to every other receiver in $O_s$, for every such $s$. For $X\subseteq R$,
\begin{equation}
\begin{split}
|E(K_{\mathrm{large}}[X])|
&\le(d+1)\sum_{s\in S_{\mathrm{large}}}|O_s\cap X|\\
&\le d(d+1)|X|.
\end{split}
\label{eq:large-degenerate-density}
\end{equation}
The first inequality bounds each sender's contribution by $d+1$ stars on $O_s\cap X$; counting an edge repeatedly only increases the bound. The second uses the participation bound at receivers.

\noindent{\bf A common coloring and the transfers.}
Let $K=K_{\mathrm{small}}\cup K_{\mathrm{large}}$. The two edge bounds give
\[
|E(K[X])|\le(d^2+3d-1)|X|\qquad(X\subseteq R).
\]
Every nonempty induced subgraph has a vertex of degree at most $2d^2+6d-2$. Lemma~\ref{lem:degeneracy} therefore gives a proper coloring $c:R\to\{1,\ldots,M_d\}$. Give each receiver the color $\gamma_r=(c(r),1)$ to use the same ordered-pair format as in Lemma~\ref{lem:sharing}. This receiver coloring is proper on $K$ and uses at most $M_d$ color pairs.

For each satisfied sender in $S_{\mathrm{small}}$, set $a_s=0$. For each unsatisfied one, select $a_s=\gamma_{\rho(s)}$. The matched receiver has a color unique in $N_B(s)$, so this adds exactly one to the even outgoing base total.

For a sender in $S_{\mathrm{large}}$, its $d+1$ candidates have distinct colors, each occurring exactly once in $O_s$. The at most $d$ exceptional receivers in $N_B(s)\setminus O_s$ can repeat at most $d$ of those colors. Hence one candidate still has a color unique among all neighbors of $s$ (see Figure \ref{fig:degenerate-candidates}). Set $q_s=0$ and select that color as $a_s$. This sender sends exactly one unit.

All senders now have odd integer outgoing totals. Incoming totals remain integral because the low-degree base contributions were integral and all corrections are integers. There are $O(d^2)$ color pairs, whose first coordinates are at most $M_d$ and whose second coordinates equal one. Each base amount has denominator at most $2d$ and numerator less than $4d$. Encoding the color coordinates separately, with a tag for the selected color, gives $O(\log(d+1))$ bits for each label.
\end{proof}

\begin{figure}[t]
\centering
\begin{minipage}[t]{.48\linewidth}
\vspace{0pt}\centering
\textbf{\small(a) Auxiliary orientation}\par\smallskip
\begin{tikzpicture}[x=.95cm,y=.95cm,font=\small,
 candidate/.style={vertex,draw=transferblue, very thick},
 orient/.style={-{Stealth[length=1.5mm]},draw=black!65}]
 \draw[draw=black!30,rounded corners=3pt, dashed] (-.5,-2.06) rectangle (.5,2.06);
 \node at (0,2.38) {$O_s$};
 \node[candidate] (r1) at (0,1.65) {$r_1$};
 \node[candidate] (r2) at (0,.55) {$r_2$};
 \node[candidate] (r3) at (0,-.55) {$r_3$};
 \node[vertex] (w) at (0,-1.65) {$w$};
 \node[vertex] (s) at (1.85,0) {$s$};
 \node[vertex] (e1) at (3.65,.65) {$e_1$};
 \node[vertex] (e2) at (3.65,-.65) {$e_2$};
 \draw[orient, color=black] (r1)--(s);
 \draw[orient, color=black] (r2)--(s);
 \draw[orient, color=black] (r3)--(s);
 \draw[orient, color=black] (w)--(s);
 \draw[orient, color=black] (s)--(e1);
 \draw[orient, color=black] (s)--(e2);
 \node[align=center,font=\footnotesize] at (3.65,1.65)
   {at most $d=2$\\exceptions};
 \node[font=\footnotesize] at (1.65,-2.43)
   {$T_s=\{r_1,r_2,r_3\}$: $d+1$ candidates};
\end{tikzpicture}
\end{minipage}\hfill
\begin{minipage}[t]{.48\linewidth}
\vspace{0pt}\centering
\textbf{\small(b) A color survives the exceptions}\par\smallskip
\begin{tikzpicture}[x=.95cm,y=.95cm,font=\small,
 candidate/.style={vertex,draw=black!75}]
 % Four colours from Paul Tol's muted qualitative palette.
 % Equal fills encode equal colours; vertex names remain inside the circles.
\definecolor{receiverblack}{HTML}{000000}
\definecolor{receiverdarkgray}{HTML}{555555}
\definecolor{receivergray}{HTML}{AAAAAA}
\definecolor{receiverwhite}{HTML}{FFFFFF}
 \draw[draw=black!30,rounded corners=3pt, dashed] (-.45,-.45) rectangle (2.05,2.05);
 \draw[draw=black!30,rounded corners=3pt, dashed] (3.1,-.45) rectangle (3.9,2.05);
 \node at (.8,2.47) {$O_s$};
 \node[align=center,font=\footnotesize] at (3.5,2.47)
   {$N_B(s)\setminus O_s$};
 \coordinate (r1) at (0,1.65);
 \coordinate (r2) at (1.6,1.65);
 \coordinate (r3) at (0,0);
 \coordinate (w) at (1.6,0);
 \draw[black] (r1)--(r2) (r1)--(r3) (r1)--(w)
   (r2)--(r3) (r2)--(w) (r3)--(w);
 \node[candidate,draw=transferblue,fill=receiverwhite, very thick] at (r1) {$r_1$};
 \node[candidate,draw=transferblue,fill=receiverdarkgray, very thick, text = white] at (r2) {$r_2$};
 \node[candidate,draw=transferblue,fill=receivergray,double,double distance=1pt, very thick]
   at (r3) {$r_3$};
 \node[vertex,draw=black,fill=receiverblack, text=white] at (w) {$w$};
 \node[vertex,draw=black,fill=receiverwhite] at (3.5,1.65) {$e_1$};
 \node[vertex,draw=black,fill=receiverdarkgray, text = white] at (3.5,0) {$e_2$};
 \node[align=center,font=\footnotesize] at (1.65,-1.45)
   {Lines are edges of $K_{\mathrm{large}}$.};
 \node[align=center,font=\footnotesize] at (1.65,-2.43)
   {The color of $r_3$ is unique in $N_B(s)$.};
\end{tikzpicture}
\end{minipage}
\caption{Ensuring a unique color when $d=2$. In (a), the orientation has at most two edges leaving $s$; all other receivers lie in $O_s$. The three blue candidates form $T_s$. In (b), each candidate is adjacent in $K_{\mathrm{large}}$ to every other receiver of $O_s$, so a proper coloring makes its color unique there. The two exceptional receivers can repeat at most two candidate colors; the double-circled candidate retains a color unique in the whole neighborhood. Arrows in (a) show only the auxiliary orientation, not the direction of transfers.}
\label{fig:degenerate-candidates}
\end{figure}

\subsection{The upper bound and certificate size}

\parityDegeneracyUpper*

\begin{proof}
Use exactly the certificate $(\lambda,\gamma,q,a)$ and verifier of Theorem~\ref{thm:upper}. In particular, $d$ is not an input or a certificate field, and the verifier does not check degeneracy.

For completeness, choose BFS layers in a connected even-order $d$-degenerate graph. Repeat the same induction from the deepest layer toward the root. At step $i$, form the bipartite graph $B_i$ from the general construction: the active vertices of $L_i$ send to their neighbors in $L_{i-1}$. Since $B_i$ is a subgraph of the input graph, it is $d$-degenerate.

Apply Lemma~\ref{lem:degenerate-sharing} to $B_i$ in place of Lemma~\ref{lem:sharing}. Active vertices send odd integer totals; inactive vertices, having received odd integer totals, send zero. Each vertex of the preceding layer receives an integer total. The same induction gives odd integer net outflow at every nonroot vertex, and conservation gives an odd integer net outflow at the root because the graph has even order. As before, each vertex gets its receiver color when the next layer is processed, and its base amount and selected color when its own layer is processed.

The layer residue takes two bits. Each receiver color or selected color pair has first coordinate at most $M_d=O(d^2)$ and second coordinate one; a selected color may also be $0$. The rational numerator and denominator are $O(d)$. The fixed encoding of the certificate therefore takes $O(\log(d+1))$ bits in total, including the tag bit for the selected color. Soundness is unchanged: transfers cancel globally for arbitrary certificates, without any degeneracy assumption.
\end{proof}

Both upper bounds use the same verifier and encoding. Consequently, on an $n$-vertex even-order graph of degeneracy at most $d$, this one verifier admits certificates of
\[
O\!\left(\min\{\log\log(n+2),\log(d+1)\}\right)
\]
bits. No scheme-selection field is needed: the prover may use either completeness construction, and neither $n$ nor $d$ is trusted input.

\subsection{A lower bound within the degeneracy promise}

The interval construction already has the appropriate degeneracy when its number of points is proportional to $d$. To obtain a lower bound under a class promise, we must also ensure that the modification used against soundness stays within that class.

\parityDegeneracyLower*

\begin{proof}
For a sufficiently large integer $d$, let $m=\max\{j\le d:j\equiv1\pmod4\}$. Use the interval incidence graph $H_m$ from Section~\ref{sec:lower}; it is connected, bipartite and its order $m(m+3)/2$ is even. Every interval vertex has degree at most $m$. Deleting all interval vertices first and then the independent point vertices gives an $m$-degeneracy ordering, so $H_m$ is $d$-degenerate.

Fix a verifier correct on connected bipartite $d$-degenerate graphs and any accepting assignment on $H_m$. As in the general lower bound, color edge $ij$ of the auxiliary $K_{m+1}$ by the certificate of $y_{ij}$. Each certificate class must be bipartite. Indeed, the argument in Lemma~\ref{lem:bipartite} turns an odd cycle of one color into interval deletions and copies that preserve every local view and change the parity of the order. The modified graph still has the same $m$ point vertices, and every interval vertex still has degree at most $m$. The same deletion ordering proves that it remains $m$-degenerate. Every connected component is thus bipartite and $d$-degenerate. An odd-order component accepted everywhere would contradict soundness even within the promised class.

If $r$ distinct certificates occur on interval vertices, Lemma~\ref{lem:bipartite-cover} gives $r\ge\lceil\log(m+1)\rceil$. If all certificates have length at most $s$, there are fewer than $2^{s+1}$ possible strings. Hence
\[
s\ge\log\log(m+1)-1=\Omega(\log\log(d+1)),
\]
where the asymptotic equality follows from $d-4<m\le d$. The argument applies to every accepting assignment and permits a different verifier for each fixed value of $d$.
\end{proof}

The witnesses have order $\Theta(d^2)$ and actual degeneracy $\Theta(d)$. For $t=(m+1)/2$, the first $t$ point vertices and the prefix interval vertices $y_{0j}$ with $t\le j\le m$ induce $K_{t,t}$. Thus their degeneracy is at least $t$ and at most $m$. The lower bound therefore concerns the structural parameter itself, rather than an unnecessarily large upper estimate for it.

\section{A lower bound on a nowhere-dense class}
\label{sec:nowhere-dense}

The bounded-degeneracy upper bound shows that many sparse classes admit constant-size certificates. We now show that the general $\Theta(\log\log n)$ bound remains tight on a fixed, subgraph-closed, bipartite nowhere-dense class. 

\parityNowhereDense*

We now give a more explicit version of
Theorem~\ref{thm:nowhere-dense}, identifying
a concrete nowhere-dense class and providing
a quantitative lower bound.

\begin{lemma}
\label{lem:nowhere-dense-explicit}
Fix $0<\alpha<1$, and let
\[
\mathcal C_\alpha=
\left\{G:\ G\text{ is bipartite and }
\operatorname{girth}(G)\ge
\bigl(\log(|V(G)|+2)\bigr)^\alpha\right\}.
\]
Then $\mathcal C_\alpha$ is subgraph-closed and nowhere dense.
For every sufficiently large even $n$, there is a connected
$n$-vertex graph $G\in\mathcal C_\alpha$ such that, for every
correct anonymous radius-one verifier for \Even\ on
$\mathcal C_\alpha$, every accepting assignment on $G$
contains a certificate of length at least
\[
\frac{1-\alpha}{3}\log\log n-O(1).
\]
Consequently, certifying \Even\ on $\mathcal C_\alpha$
has local complexity $\Theta(\log\log n)$.
\end{lemma}

We return to the interval idea from Section~\ref{sec:lower}, but use identities between color counts instead of identities between neighborhoods. This lets us change parity by splitting a single vertex, an operation that cannot create a shorter cycle.

For a certificate assignment $c$ using at most $r$ distinct strings, identify these strings with $[r]:=\{1,\ldots,r\}$ and define the \emph{profile}
\[
\mathbf m_c(v)=\bigl(|\{w\in N(v):c(w)=a\}|\bigr)_{a\in[r]}.
\]
Thus the local view at $v$ is determined by $c(v)$ and $\mathbf m_c(v)$. The same formula defines $\mathbf m_c(v)$ when $c$ is defined only on a set containing $N(v)$. These colorings need not be proper.

\begin{lemma}
\label{lem:profile-splitting}
Let $c$ be a certificate assignment on a bipartite graph $G$. Suppose that distinct vertices $u,v,w$ have the same certificate and
\[
\mathbf m_c(v)=\mathbf m_c(u)+\mathbf m_c(w).
\]
There are a simple bipartite graph $G'$ of order $|V(G)|+1$ and girth at least that of $G$, and a certificate assignment $c'$ on $G'$, such that every local view in $(G',c')$ occurs in $(G,c)$.
\end{lemma}

\begin{proof}
Replace $v$ by two vertices $v_1,v_2$, both carrying $c(v)$. Partition its neighbors within each certificate class so that $v_1$ has profile $\mathbf m_c(u)$ and $v_2$ has profile $\mathbf m_c(w)$. This is possible by the assumed identity. The new vertices imitate $u$ and $w$. Every retained vertex keeps its certificate and the number of its neighbors with each certificate, since any former neighbor of $v$ is adjacent to exactly one replacement carrying $c(v)$.

Put $v_1,v_2$ on the same side of the bipartition as $v$. Identifying them back into $v$ maps any new cycle to a closed trail in $G$: no edge repeats, since each old edge was assigned to exactly one replacement. The trail contains a cycle of no greater length, so girth does not decrease. The order increases by one.
\end{proof}

For example, profiles $(1,1)$, $(1,2)$, and $(2,3)$ allow the last vertex to be split into two vertices imitating the first two, provided all three carry the same certificate. If splitting disconnects the graph, every component still retains all its local views.

The next lemma supplies such a triple under every small certificate palette. Its property survives adding leaves at a vertex of $P$, because none of the old neighborhoods on the $Y$ side changes.

\begin{lemma}
\label{lem:profile-core}
There is an absolute constant $C\ge1$ such that, for all integers $r\ge2$ and $g\ge4$, there is a connected bipartite graph $H=(P\cup Y,E)$ of girth at least $g$ and order at most $2^{Cgr^2\log r}$ such that every coloring $c\colon V(H)\to[r]$ has distinct vertices $u,v,w\in Y$ with the same color and nonzero profiles satisfying
\[
\mathbf m_c(v)=\mathbf m_c(u)+\mathbf m_c(w).
\]
\end{lemma}

Here is the idea of the proof, deferred to Appendix~\ref{app:profile-core}. An ordered sequence of points gives interval profiles satisfying the exact identity $\mathbf z_{ik}=\mathbf z_{ij}+\mathbf z_{jk}$ for $i<j<k$. For each coloring of $P$, a counting argument finds a sequence whose interval profiles all have a reasonable probability of appearing in random subsets. Sampling supplies many vertices with each of these profiles, simultaneously for all colorings, so deleting one set vertex from each short cycle still leaves every profile available. Select a representative for each interval and color its endpoint pair by the representative's certificate. A monochromatic triangle gives the required three equally certified vertices. No approximation of color frequencies is needed. Figure~\ref{fig:profile-splitting} illustrates how the triangle gives the split.

\begin{figure}[t]
\centering
\begingroup
\definecolor{profilewhite}{HTML}{FFFFFF}
\definecolor{profiledarkgray}{HTML}{333333}
\tikzset{
 profilepoint/.style={circle,draw=black!75,minimum size=5mm,inner sep=1pt},
 profileone/.style={profilepoint,fill=profiledarkgray, },
 profiletwo/.style={profilepoint,fill=profilewhite},
 profilecenter/.style={profileone,minimum size=6mm,thick},
 profileedge/.style={draw=black!75,line width=.6pt}
}
\begin{minipage}[t]{.32\linewidth}
\vspace{0pt}\centering
\textbf{\small(a) Auxiliary triangle}\par\smallskip
\begin{tikzpicture}[x=.93cm,y=.9cm,font=\small]
 \path[use as bounding box] (-.4,-.45) rectangle (3.6,2.8);
 \node[vertex,minimum size=5mm] (i) at (0,.2) {$i$};
 \node[vertex,minimum size=5mm] (j) at (1.6,2.25) {$j$};
 \node[vertex,minimum size=5mm] (k) at (3.2,.2) {$k$};
 \draw[transferblue,color=black,thick] (i)--
   node[sloped,above=2pt,font=\small] {$u:\mathbf a$} (j);
 \draw[transferblue,color=black,thick] (j)--
   node[sloped,above=2pt,font=\small] {$w:\mathbf b$} (k);
 \draw[transferblue,color=black,thick] (i)--
   node[below=3pt,font=\small] {$v:\mathbf a+\mathbf b$} (k);
 \node[font=\scriptsize,align=center] at (1.6,.95)
   {common\\certificate $1$};
\end{tikzpicture}
\end{minipage}\hfill
\begin{minipage}[t]{.32\linewidth}
\vspace{0pt}\centering
\textbf{\small(b) Before the split}\par\smallskip
\begin{tikzpicture}[x=.93cm,y=.9cm,font=\small]
 \path[use as bounding box] (-.4,-.45) rectangle (3.6,2.8);
 \coordinate (v) at (1.6,2.05);
 \coordinate (x1) at (0,.2);
 \coordinate (x2) at (.8,.2);
 \coordinate (x3) at (1.6,.2);
 \coordinate (x4) at (2.4,.2);
 \coordinate (x5) at (3.2,.2);
 \draw[profileedge] (v)--(x1) (v)--(x2) (v)--(x3)
   (v)--(x4) (v)--(x5);
 \node[profilecenter,label=above:$v$, text=white] at (v) {$1$};
 \node[profileone,label=below:$x_1$, text=white] at (x1) {$1$};
 \node[profiletwo,label=below:$x_2$] at (x2) {$2$};
 \node[profileone,label=below:$x_3$, text=white] at (x3) {$1$};
 \node[profiletwo,label=below:$x_4$] at (x4) {$2$};
 \node[profiletwo,label=below:$x_5$] at (x5) {$2$};
\end{tikzpicture}
\end{minipage}\hfill
\begin{minipage}[t]{.32\linewidth}
\vspace{0pt}\centering
\textbf{\small(c) After the split}\par\smallskip
\begin{tikzpicture}[x=.93cm,y=.9cm,font=\small]
 \path[use as bounding box] (-.4,-.45) rectangle (3.6,2.8);
 \coordinate (v1) at (.4,2.05);
 \coordinate (v2) at (2.4,2.05);
 \coordinate (x1) at (0,.2);
 \coordinate (x2) at (.8,.2);
 \coordinate (x3) at (1.6,.2);
 \coordinate (x4) at (2.4,.2);
 \coordinate (x5) at (3.2,.2);
 \draw[profileedge] (v1)--(x1) (v1)--(x2)
   (v2)--(x3) (v2)--(x4) (v2)--(x5);
 \node[profilecenter,label=above:$v_1$, text=white] at (v1) {$1$};
 \node[profilecenter,label=above:$v_2$, text=white] at (v2) {$1$};
 \node[profileone,label=below:$x_1$, text=white] at (x1) {$1$};
 \node[profiletwo,label=below:$x_2$] at (x2) {$2$};
 \node[profileone,label=below:$x_3$, text=white] at (x3) {$1$};
 \node[profiletwo,label=below:$x_4$] at (x4) {$2$};
 \node[profiletwo,label=below:$x_5$] at (x5) {$2$};
\end{tikzpicture}
\end{minipage}
\[
\mathbf a=(1,1),\qquad \mathbf b=(1,2),\qquad
\mathbf a+\mathbf b=(2,3),\qquad |V(G')|=|V(G)|+1.
\]
\endgroup
\caption{A monochromatic triangle yields a split. The auxiliary edges $ij,jk,ik$ in (a) have representatives $u,w,v$ with certificate $1$ and profiles $\mathbf a,\mathbf b,\mathbf a+\mathbf b$. Panels (b,c) show $v$'s actual neighborhood: $v_1$ and $v_2$ imitate $u$ and $w$ by receiving the two indicated groups. Every old neighbor still sees one vertex with certificate $1$. Fills and numeric labels encode certificates. The templates are matched by color counts; their actual neighbors may differ. Unshown vertices and edges are unchanged.}
\label{fig:profile-splitting}
\end{figure}

\begin{proof}[Proof of \Cref{lem:nowhere-dense-explicit}]
Taking a subgraph does not decrease girth or increase order, so $\mathcal C_\alpha$ is subgraph-closed. A depth-$\rho$ minor model of $K_3$ yields a cycle of length at most $6\rho+3$: join the endpoints of its three connecting edges inside their branch sets by paths of length at most $2\rho$ each. These paths have disjoint interiors, since the branch sets are disjoint. For fixed $\rho$, sufficiently large members of $\mathcal C_\alpha$ therefore exclude this minor. The remaining members have bounded order and exclude a sufficiently large complete minor. This proves nowhere denseness.

Fix a sufficiently large even $n$, write $L=\log n$, and choose
\[
g=\left\lceil\bigl(\log(n+3)\bigr)^\alpha\right\rceil,
\qquad
r=\left\lfloor L^{(1-\alpha)/3}\right\rfloor.
\]
Then $g\ge4$ and $r\ge2$. Since $g=O(L^\alpha)$, Lemma~\ref{lem:profile-core} gives a graph $H$ with
\[
\log|V(H)|=O(gr^2\log r)
=O\!\left(L^{(2+\alpha)/3}\log L\right)=o(L).
\]
Here the last equality uses $\alpha<1$. Thus $|V(H)|\le\sqrt n$ for large $n$. Add leaves at one vertex of $P$ until the order is exactly $n$. The resulting connected graph $G_n$ lies in $\mathcal C_\alpha$ and retains the lemma's profile property.

Suppose an accepting assignment $c$ on $G_n$ uses at most $r$ distinct certificates. Its restriction to $H$ gives equally certified vertices $u,v,w\in Y$ with $\mathbf m_c(v)=\mathbf m_c(u)+\mathbf m_c(w)$.
These are also their profiles in $G_n$, since adding leaves at $P$ did not change their neighborhoods. Lemma~\ref{lem:profile-splitting} produces an accepted bipartite graph of odd order $n+1$ and girth at least $g$. It has an odd-order component $F$. This component is a connected accepted no-instance in the same promised class, since
\[
\operatorname{girth}(F)\ge g\ge
\bigl(\log(|V(F)|+2)\bigr)^\alpha.
\]
This contradicts soundness on $\mathcal C_\alpha$.

Every accepting assignment thus uses more than $r$ distinct strings. If their lengths are at most $s$, then $r+1\le2^{s+1}-1$. Since $r+1>L^{(1-\alpha)/3}$,
\[
s\ge\log(r+1)-1
\ge\frac{1-\alpha}{3}\log\log n-1.
\]
The upper bound follows from Theorem~\ref{thm:upper}.
\end{proof}

Together with Theorem~\ref{thm:upper},
Lemma~\ref{lem:nowhere-dense-explicit}
implies Theorem~\ref{thm:nowhere-dense}.

\section*{AI disclosure}
The writing of this paper and the derivation of its results were assisted by AI agents. We used Claude Code (Anthropic) with the model Claude Opus 5.5, and Codex (OpenAI) with the model GPT-6.1 Sol. The authors reviewed all text and results and take full responsibility for them.

\bibliography{references}

@misc{BFVZ,
  author        = {Nicolas Bousquet and Laurent Feuilloley and Jorge Valenzuela and S{\'e}bastien Zeitoun},
  title         = {The Local Complexity of Certifying Parity},
  year          = {2026},
  eprint        = {2606.04934},
  archivePrefix = {arXiv},
  primaryClass  = {cs.DC},
  note          = {DISC 2026; full version}
}

@article{KKP10,
  author  = {Amos Korman and Shay Kutten and David Peleg},
  title   = {Proof Labeling Schemes},
  journal = {Distributed Computing},
  volume  = {22},
  number  = {4},
  pages   = {215--233},
  year    = {2010},
  doi     = {10.1007/s00446-010-0095-3}
}

@article{GS16,
  author  = {Mika G{\"o}{\"o}s and Jukka Suomela},
  title   = {Locally Checkable Proofs in Distributed Computing},
  journal = {Theory of Computing},
  volume  = {12},
  number  = {19},
  pages   = {1--33},
  year    = {2016},
  doi     = {10.4086/toc.2016.v012a019},
  url     = {https://theoryofcomputing.org/articles/v012a019/}
}

@article{FKP13,
  author  = {Pierre Fraigniaud and Amos Korman and David Peleg},
  title   = {Towards a Complexity Theory for Local Distributed Computing},
  journal = {Journal of the ACM},
  volume  = {60},
  number  = {5},
  pages   = {35:1--35:26},
  year    = {2013},
  doi     = {10.1145/2499228}
}

@article{F21,
  author  = {Laurent Feuilloley},
  title   = {Introduction to Local Certification},
  journal = {Discrete Mathematics \& Theoretical Computer Science},
  volume  = {23},
  number  = {3},
  year    = {2021},
  doi     = {10.46298/dmtcs.6280}
}

@inproceedings{BFZ25,
  author    = {Nicolas Bousquet and Laurent Feuilloley and S{\'e}bastien Zeitoun},
  title     = {Complexity Landscape for Local Certification},
  booktitle = {39th International Symposium on Distributed Computing (DISC 2025)},
  series    = {Leibniz International Proceedings in Informatics (LIPIcs)},
  volume    = {356},
  pages     = {18:1--18:21},
  publisher = {Schloss Dagstuhl -- Leibniz-Zentrum f{\"u}r Informatik},
  year      = {2025},
  doi       = {10.4230/LIPIcs.DISC.2025.18}
}

@article{H23,
  author  = {Robert Hickingbotham},
  title   = {Odd Colourings, Conflict-Free Colourings and Strong Colouring Numbers},
  journal = {Australasian Journal of Combinatorics},
  volume  = {87},
  number  = {1},
  pages   = {160--164},
  year    = {2023},
  url     = {https://ajc.maths.uq.edu.au/pdf/87/ajc_v87_p160.pdf}
}

@article{Planar21,
  author  = {Laurent Feuilloley and Pierre Fraigniaud and Pedro Montealegre and Ivan Rapaport and {\'E}ric R{\'e}mila and Ioan Todinca},
  title   = {Compact Distributed Certification of Planar Graphs},
  journal = {Algorithmica},
  volume  = {83},
  number  = {7},
  pages   = {2215--2244},
  year    = {2021},
  doi     = {10.1007/s00453-021-00823-w}
}

@inproceedings{BFP22,
  author    = {Laurent Feuilloley and Nicolas Bousquet and Th{\'e}o Pierron},
  title     = {What Can Be Certified Compactly? Compact Local Certification of {MSO} Properties in Tree-Like Graphs},
  booktitle = {Proceedings of the ACM Symposium on Principles of Distributed Computing (PODC 2022)},
  pages     = {131--140},
  publisher = {ACM},
  year      = {2022},
  doi       = {10.1145/3519270.3538416}
}

@article{FMRT24,
  author  = {Pierre Fraigniaud and Pedro Montealegre and Ivan Rapaport and Ioan Todinca},
  title   = {A Meta-Theorem for Distributed Certification},
  journal = {Algorithmica},
  volume  = {86},
  number  = {2},
  pages   = {585--612},
  year    = {2024},
  doi     = {10.1007/s00453-023-01185-1}
}

@inproceedings{CKM25,
  author    = {Linda Cook and Eun Jung Kim and Tom{\'a}{\v s} Masa{\v r}{\'i}k},
  title     = {A Tight Meta-Theorem for {LOCAL} Certification of {MSO}$_2$ Properties Within Bounded Treewidth Graphs},
  booktitle = {Proceedings of the ACM Symposium on Principles of Distributed Computing (PODC 2025)},
  pages     = {110--120},
  publisher = {ACM},
  year      = {2025},
  doi       = {10.1145/3732772.3733562}
}

@article{Hall35,
  author  = {Philip Hall},
  title   = {On Representatives of Subsets},
  journal = {Journal of the London Mathematical Society},
  volume  = {s1-10},
  number  = {1},
  pages   = {26--30},
  year    = {1935},
  doi     = {10.1112/jlms/s1-10.37.26}
}

@article{SW68,
  author  = {George Szekeres and Herbert S. Wilf},
  title   = {An Inequality for the Chromatic Number of a Graph},
  journal = {Journal of Combinatorial Theory},
  volume  = {4},
  number  = {1},
  pages   = {1--3},
  year    = {1968},
  doi     = {10.1016/S0021-9800(68)80081-X}
}

@article{HHM77,
  author  = {Frank Harary and Derbiau Hsu and Zevi Miller},
  title   = {The Biparticity of a Graph},
  journal = {Journal of Graph Theory},
  volume  = {1},
  number  = {2},
  pages   = {131--133},
  year    = {1977},
  doi     = {10.1002/jgt.3190010208}
}

@article{DOP24,
  author  = {Tianjiao Dai and Qiancheng Ouyang and Fran{\c c}ois Pirot},
  title   = {New Bounds for Odd Colourings of Graphs},
  journal = {The Electronic Journal of Combinatorics},
  volume  = {31},
  number  = {4},
  pages   = {P4.57},
  year    = {2024},
  doi     = {10.37236/12110}
}

@article{CKP13,
  author  = {Panagiotis Cheilaris and Bal{\'a}zs Keszegh and D{\"o}m{\"o}t{\"o}r P{\'a}lv{\"o}lgyi},
  title   = {Unique-Maximum and Conflict-Free Coloring for Hypergraphs and Tree Graphs},
  journal = {SIAM Journal on Discrete Mathematics},
  volume  = {27},
  number  = {4},
  pages   = {1775--1787},
  year    = {2013},
  doi     = {10.1137/120880471}
}

@article{ELRS03,
  author  = {Guy Even and Zvi Lotker and Dana Ron and Shakhar Smorodinsky},
  title   = {Conflict-Free Colorings of Simple Geometric Regions with Applications to Frequency Assignment in Cellular Networks},
  journal = {SIAM Journal on Computing},
  volume  = {33},
  number  = {1},
  pages   = {94--136},
  year    = {2003},
  doi     = {10.1137/S0097539702431840}
}

@article{PT09,
  author  = {J{\'a}nos Pach and G{\'a}bor Tardos},
  title   = {Conflict-Free Colourings of Graphs and Hypergraphs},
  journal = {Combinatorics, Probability and Computing},
  volume  = {18},
  number  = {5},
  pages   = {819--834},
  year    = {2009},
  doi     = {10.1017/S0963548309990290}
}

@book{NODM12,
  author    = {Jaroslav Ne{\v s}et{\v r}il and Patrice {Ossona de Mendez}},
  title     = {Sparsity: Graphs, Structures, and Algorithms},
  series    = {Algorithms and Combinatorics},
  volume    = {28},
  publisher = {Springer},
  year      = {2012},
  doi       = {10.1007/978-3-642-27875-4}
}

@misc{FSS25,
  author = {Laurent Feuilloley and Josef Erik Sedl{\'a}{\v c}ek and Martin Sl{\'a}vik},
  title = {Proving there is a leader without naming it},
  year = {2025},
  eprint = {2511.15491},
  archivePrefix = {arXiv},
  primaryClass = {cs.DC},
  url = {https://arxiv.org/abs/2511.15491}
}

@article{PSP22,
  author  = {Boaz Patt-Shamir and Mor Perry},
  title   = {Proof-labeling schemes: Broadcast, unicast and in between},
  journal = {Theoretical Computer Science},
  volume  = {923},
  pages   = {179--195},
  year    = {2022},
  doi     = {10.1016/j.tcs.2022.05.006}
}

@misc{CLT26,
  author = {Yi-Jun Chang and Yi-Xuan Lee and Meng-Tsung Tsai},
  title = {Local Certification of Vertex and Edge Connectivity},
  year = {2026},
  eprint = {2607.13677},
  archivePrefix = {arXiv},
  primaryClass = {cs.DC},
  url = {https://arxiv.org/abs/2607.13677}
}

\appendix

\section{Proof of \texorpdfstring{Lemma~\ref{lem:profile-core}}{the profile lemma}}
\label{app:profile-core}
\begin{proof}
We need three equally colored vertices whose profiles satisfy an additive identity. Intervals provide the identity; a monochromatic triangle will provide the equal colors. The random construction only has to realize enough interval profiles to make this argument survive the removal of short cycles. Throughout the proof, $\ln$ denotes the natural logarithm.

Put
\[
D=3r!,\qquad \eta=(D+1)^{-(r+3)},\qquad
\Lambda=\left\lceil16\eta^{-1}\ln(2rD)\right\rceil,
\qquad N=(16\Lambda D^3)^g.
\]
Let $P$ be a fixed set of $N$ points.

\paragraph*{Every coloring has a sequence of likely interval profiles.}
Fix $c\colon P\to[r]$. For $1\le j\le D$, let $q_j(\mathbf z)$ be the probability that a uniform $j$-element subset of $P$ has profile $\mathbf z$. There are at most $(D+1)^r$ possible profiles, since each coordinate is an integer between $0$ and $D$. Consequently, a random $j$-subset has a profile $\mathbf z$ with $q_j(\mathbf z)<\eta$ with probability at most $(D+1)^r\eta$.

Now choose a uniform ordered sequence $x_1,\ldots,x_D$ of distinct points. For each interval $(i,j]$ of positions, its points form a uniform $(j-i)$-subset of $P$. A union bound over fewer than $D^2$ intervals gives
\[
\Pr\bigl(\text{some interval has a profile of probability }<\eta\bigr)
\le D^2(D+1)^r\eta=\frac{D^2}{(D+1)^3}<1.
\]
Thus we can fix a sequence for which every interval profile $\mathbf z_{ij}=\sum_{i<k\le j}\mathbf e_{c(x_k)}$, with $0\le i<j\le D$, has probability at least $\eta$ under sampling a subset of its own size. Here $\mathbf e_a$ is the $a$th unit vector. These nonzero profiles satisfy
\begin{equation}
\mathbf z_{ik}=\mathbf z_{ij}+\mathbf z_{jk}\qquad(i<j<k).
\label{eq:interval-profile-addition}
\end{equation}
Fix one such sequence for each coloring of $P$, before making the random graph. The sequence may depend on the coloring; the graph will work for all these choices simultaneously.

\paragraph*{Realize the profiles, then remove short cycles.}
For every $1\le j\le D$, independently sample $\Lambda N$ uniform $j$-element subsets of $P$. Represent each sample by a new vertex adjacent to its elements. Repeated subsets give distinct vertices, so this incidence graph is simple and bipartite.

Fix a coloring of $P$ and one interval profile from its fixed sequence. The number $X$ of samples realizing that profile is binomial, with mean $\mu\ge\Lambda N\eta\ge16N\ln(2rD)$. Exponential Markov gives the elementary tail estimate
\[
\Pr(X<N)\le2^N\mathbb E[2^{-X}]
\le\exp(N\ln2-\mu/2)
\le(2rD)^{-2N}.
\]
Indeed, for a binomial variable with $M$ trials and success probability $q$, $\mathbb E[2^{-X}]=(1-q/2)^M\le e^{-Mq/2}$. There are $r^N$ colorings and fewer than $D^2$ fixed interval targets per coloring. Hence
\[
\Pr(\text{some target has fewer than }N\text{ copies})
\le r^N D^2(2rD)^{-2N}
\le(4r)^{-N}<\frac14.
\]

Let $Z$ count cycles of length less than $g$. A cycle of length $2\ell$ uses $\ell$ distinct points and $\ell$ distinct set vertices. There are at most $N^\ell(D\Lambda N)^\ell$ ordered choices. A sampled set contains two specified distinct points with probability at most $2D^2/N^2$, and different set vertices were sampled independently. Therefore
\[
\mathbb E Z
\le\sum_{2\le\ell<g/2}(2\Lambda D^3)^\ell
\le g(2\Lambda D^3)^g
=\frac{g}{8^g}N\le\frac N8.
\]
By Markov, $Z\le N/2$ with probability at least $3/4$. Some realization therefore has both properties. Delete one set vertex from each short cycle. At most $N/2$ vertices are deleted, leaving at least $N/2$ copies of every fixed interval target for every coloring of $P$. All surviving set vertices keep their neighborhoods.

Every component contains a point. Connect the components along a tree by adding paths of length two between points in different components, with fresh internal vertices on the set side. The added edges are bridges. Thus girth is at least $g$, and the surviving original set vertices still keep their profiles. The final graph is connected and has at most $(D\Lambda+2)N$ vertices.

\paragraph*{A monochromatic triangle gives the split.}
Every edge coloring of $K_{3r!}$ with $r$ colors has a monochromatic triangle; this follows by induction on $r$. It is immediate for $r=1$. For $r\ge2$, some vertex has at least $3(r-1)!$ neighbors joined to it in one color. Either that color occurs on an edge between those neighbors, completing a triangle, or their complete graph uses at most $r-1$ colors and induction applies.

Take any coloring of the whole constructed graph with at most $r$ colors, and use the sequence fixed for its restriction to $P$. For every pair $0\le i<j\le D$, choose a surviving original set vertex $y_{ij}$ of profile $\mathbf z_{ij}$. The representatives can be chosen distinct: fewer than $D^2$ are needed, while every target has at least $N/2>D^2$ available vertices. Color edge $ij$ of the complete graph on $\{0,\ldots,D\}$ by $c(y_{ij})$. The preceding Ramsey bound gives a monochromatic triangle $i<j<k$. Its three representatives have the same color, and~\eqref{eq:interval-profile-addition} gives $\mathbf m_c(y_{ik})=\mathbf m_c(y_{ij})+\mathbf m_c(y_{jk})$. This is the required triple.

Finally,
\[
\log D=O(r\log r),\qquad
\log\Lambda=O(r^2\log r),\qquad
\log N=O(gr^2\log r).
\]
The order bound $(D\Lambda+2)N$ is therefore at most $2^{Cgr^2\log r}$ for an absolute constant $C\ge1$.
\end{proof}

\end{document}